\documentstyle[11pt]{article}

\newtheorem{example}{Example}[section]
\newtheorem{lemma}{Lemma}[section]
\newtheorem{theorem}{Theorem}[section]
\newtheorem{definition}{Definition}[section]

\begin{document}

\begin{center}
{\huge \bf A  Development Calculus \\[3mm]
for   Specifications }
\end{center}

\begin{center}
 Wei \hspace{5mm} LI

\end{center}

\begin{center}
liwei@nslde.buaa.edu.cn\\
 State Key  Laboratory of Software Development Environment\\
Beihang University \\
 100083 Beijing, P.R. China
\end{center}

\vspace{5mm}

\noindent {\bf Abstract}

A first order inference system, named R-calculus, is defined to
develop the specifications. This system intends to eliminate the
laws which is not consistent with user's requirements. The
R-calculus consists of the structural rules, an axiom, a cut rule,
and the rules for logical connectives. Some examples are given to
demonstrate the usage of the R-calculus. Furthermore, the
properties regarding reachability and completeness of the
R-calculus are formally defined and proved.

\vspace{3mm}

\noindent {\bf Keywords: Specification, Revision, Necessary
premise, R-calculus, R-condition}

\section{ Motivation}

During the development of the specifications, two situations are
encountered commonly: First, the specification is not consistent;
as a result, a program to satisfy it does not exist. Second, the
specification is consistent, but users and designers refuse to
accept it since the running results of the generated program does
not meet their requirements. Under both circumstances, the
specification should be redesigned. If it was indeed the case, the
elimination of inconsistent laws and the introduction of new laws
become very important; this paper will address this concerns
principally. For example, consider the following specification:

\[
\Gamma \equiv \{ A, A \supset B, B \supset C, E\supset F\}
\]
where $A$, $B$, $C$, $E$, and $E$ denote some equations. It is
obviously that $\Gamma \vdash C$ holds.  Users, however, reject
$C$ and prefer $\neg C $. In this case, we have to redesign the
specification.

The purpose of this paper, therefore, concerns principally to
build an inference system, named R-calculus to revise the
specifications. The nature of R-calculus differentiates it from
many well known inference systems of specifications. The purpose
of the latter aims to deduce the correct programs to satisfy a
given specification[10,11,12,13]. However, R-calculus emphasizes
on revising the specifications, and eliminating the laws which are
not consistent with user's requirements.

The R-calculus is indeed a transition system which consists of the
structural rules, an axiom, an cut rule, and the rules of logical
connectives. Some examples are given to demonstrate how the
R-calculus can be used to develop specifications. Furthermore, the
properties of the R-calculus, such as reachability and
completeness, are formally defined and proved.

\section{The Necessary Premise}

In order to avoid the syntactical details, in this paper,
the first order languages are chosen to be the specification languages
[1]. Briefly, a first order language $\cal L$ has two sets of
symbol strings. They are the set of terms and the set of formulas.
The set of terms is defined on the set of variable symbols {\bf V}
ranged over by $x,y,z \cdots $, the set of function symbols {\bf
F} ranged over by $f,g,h \cdots $, and the set of constants
symbols {\bf C} ranged over by $a,b,c \cdots$, and it is defined
inductively as below:
\[
t~ ::=~ c~ | ~x~ |~ f(t_1, t_2, \cdots, t_n)
\]
 The set of formulas are defined on the set of predicates $\bf P$
 ranged over by $P,Q, R, \cdots$,
 and the set of logical connectives including:
 $\neg$, $\land$, $\lor$, $\supset$, $\forall$,  $\exists$,
 and it is defined inductively as the following,
\[
A~::=~P(t_1, t_2, \cdots, t_n)~|~ \neg A~|~A \land B~|~A\lor B~|~
A \supset B~| \forall x.A ~|~ \exists x. A
\]

In this paper, $\Gamma$ is used to denote a formal theory,
which is a finite set of formulas. $Th(\Gamma)$ denotes
the set of all logical consequences of $\Gamma$.
$\Gamma \vdash A$ is called a sequent, where
 $A$ is a logical consequence of $\Gamma$ [1,2],
$\vdash $ is the deductive relation.
A Gentzen style inference system, such as {\bf G} system [1] is
employed for the logical analysis of the specifications.
Each inference rule of LK is described by a fraction of number of sequents.
A proof tree {\bf T} of the sequent $\Gamma \vdash A$ is a finite tree structure,
where every node of {\bf T} is a sequent, the node and its direct sons
forms an application of an inference rule of the {\bf G} system,
the root of {\bf T} is $\Gamma \vdash A$, and every leaf of {\bf T} is
an axiom.

\begin{definition}
\rm
 Necessary premise

Let $\Gamma \vdash A$, and {\bf T} be its proof tree. Let $P$, $Q$,
and $R$ be formulas in {\bf T}. $P$ is
premise of $Q$, if and only if the following items hold:

\begin{enumerate}

\item If $\Gamma',P \vdash P$ is a leaf of {\bf T},
then the $P$ on the left hand side of $\vdash$ is the
premise of the $P$ on the right hand.

\item If the node of {\bf T} is an application of a right rule,
$Q$ is one of $A\land B$, $A \lor B$,$A \supset B$, $\neg A$,
$\forall x.A$, and $\exists x.A$, which is a principal formula [1]
in the denominator of the inference rules, and $P$ is one of
$A$,$B$,$A[t/x]$, and $A[y/x]$, which is a side formula [1] in the
numerator of the corresponding inference rule,then $P$ is
the premise of $Q$.

\item If the node of {\bf T} is an application of a left rule, and
$Q'$ is one of $A\land B$, $A \lor B$,$A \supset B$, $\neg A$,
$\forall x.A$, and $\exists x.A$,
 which is a principal formula in the denominator of the inference rules,
then $Q'$ is the premise of $A$,$B$, $A[t/x]$, and
$A[y/x]$, which is a side formula in the numerator of the
corresponding inference rule.
Every one of $A$,$B$, $A[t/x]$, and $A[y/x]$ is
  necessary premise of the formula on the right hand side of
$\vdash$ in the denominator.

\item If $P$ is $Q$'s premise, and $Q$ is $R$'s necessary,
then $P$ is $R$'s necessary premise.

\end{enumerate}

Let ${\cal P}_{\mbox{\small \bf T}}(\Gamma, A)$ be the set of premise of $A$ in the proof tree {\bf T}. If $P \in
\Gamma$ holds and $P$ is the premise of $A$ in {\bf T}, precisely, $P \in \Gamma \cap {\cal P}_{\mbox{\small \bf T}}(\Gamma, A)$,
$P$ is the {\b necessary} premise of $A$ in {\bf T}, which can be
written as $P \mapsto_{\mbox{\small \bf T}} A$. \hfill $Box$

\end{definition}

According to definition 2.1, for any given $\Gamma \vdash A$, the necessary premise of $A$ depends on the proof tree {\bf T}. However, whenever $\Gamma \vdash A$ holds, its proof tree exists.
Thus, for the reason to simplify the writings, sometimes the tree {\bf T} is omitted from $\mapsto $ and the related notation will be written as $P \mapsto A$
when there does not exist confusion in the context.

\begin{example}
\rm  $\land$- right rule:
\[
\frac{\Gamma \vdash A~~~\Gamma \vdash B}{\Gamma \vdash A\land B}
\]
$A$ and $B$ are the necessary premise of $A \land B$.

\end{example}

\begin{example}\

\rm
 Consider the sequent: $C,A,\forall x ( A \supset
B(x))\vdash \exists x B(x)$. Its proof tree is the following:

\vspace{3mm}

\noindent \mbox{}\hspace{-2mm} {\small $
\begin{array}{c}
\underline{C,A^{\ast 4},(\forall x(A\supset B(x)))^{\ast 2},(A
\supset B[t/x] )^{\ast 2} \vdash A^{\ast 3}\ \ \ \ \ C,A,(\forall
x(A\supset B(x)))^{\ast 2},(B[t/x])^{\ast 3}
\vdash B[t/x]^{\ast 1}}\\
\hspace{2.5cm}\underline{ C,A,(\forall x (A\supset B(x)))^{\ast
2},(A\supset B[t/x] )^{\ast 2}
\vdash B[t/x]^{\ast 1}} \\
\hspace{3cm}\underline{ C,A,(\forall x (A\supset B(x)))^{\ast 2 } \vdash B[t/x]^{\ast 1}} \\
\hspace{3.5cm} C,A,\forall x(A\supset B(x)) \vdash \exists x B(x)
\end{array}
$ }

\vspace{2mm}

The first node is an application of the $\exists$-right rule.
$B[t/x]$ is premise of $\exists x B(x)$. We use
superscript $\ast$ to denote the premise, and use
number $1$ to denote the first node. The second node is an
application of the $\forall$-left rule. According to
definition 2.1, $(\forall x (A\supset B(x)))^{\ast 2}$ on the left
hand of $\vdash $ in the denominator of the node 2 is the
premise of $(A \supset B[t/x])^{\ast 2}$ on the left hand of
$\vdash$ in the numerator, and $(A\supset B[t/x])^{\ast 2}$ is
also the premise of $ B[t/x]^{\ast 1}$. The third node
of the proof three is an application of $\supset$-left rule.
According to definition 2.1, $(A\supset B[t/x])^{\ast 2}$ is
the premise of $A^{\ast 3}$ on the right hand of
$\vdash$ of the first sequent in the numerator, and is also the
premise of $ B[t/x]^{\ast 3}$ on the left hand of
$\vdash$ of the second t in the numerator. $A^{\ast 3}$ and
$B[t/x]^{\ast 3}$ are the premise of $ B[t/x]^{\ast 1}$
on the right hand of $\vdash$ in the denominator of the node 3.
The forth node is an application of the axiom. $A^{\ast 4}$ on the
left hand of $\vdash$ is the premise of $A^{\ast 3}$ on
the right hand. The fifth node is also an application of the
axiom. $ B[t/x]^{\ast 3}$ on the left hand of $\vdash$ is the
premise of $ B[t/x]^{\ast 1}$ on the right. Thus, the
set of numbers of  the premise of the proof three is
\[
\{ B[t/x] ,\forall x (A\supset B(x)),A\supset B[t/x],A \}
\]
According to the definition 2.1, the necessary premise of $\exists x.B(x)$
of the sequent $C,A,\forall x( A\supset B(x))\vdash \exists x.B(x)$ is:
\[
\{A,\forall x (A\supset B(x))\}
\]
\hfill $\Box$

\end{example}

\begin{lemma}
\rm

Let $\Gamma \vdash A$ and {\bf T} be its proof tree.
The set ${\cal P}(\Gamma, A)$ is decidable.
\end{lemma}

\noindent {\bf Proof.} According the definition of the
necessary premise, an algorithm can be designed in the following way:
Its input is the proof tree, and its output is the set ${\cal P}(\Gamma, A)$.
The algorithm computes the premise from the root of {\bf T} to the leaves of
{\bf T} as shown in the above example2.2.
Since the proof tree {\bf T} is finite, the algorithm will be halt. \hfill $\Box$

In this paper, the finite formal theories of $\cal L$ are used to describe the
specifications.

\begin{definition}

Specification

\rm

A finite consistent set $\Gamma$ of the sentences is called a
specification. The sentences contained in $\Gamma$ are called the
laws of the specification.

\end{definition}

 We assume that two sentences $P$ and $Q$ are the same sentence
 if and only if $P \equiv Q$ (that is ($P \supset Q) \land (Q\supset P$)
 is a tautology).

A model ${\bf M}$ is a pair $<M,I>$, where $M$ is a non
empty set and it is called domain,
 $I$ is a map and it is called interpretation.
The form ${\bf M} \models A$ means that
 for the given domain $M$ and the interpretation $I$, $A$ is true in $M$.
 ${\bf M} \models \Gamma $ mean that for every $A \in \Gamma$, ${\bf M} \models A$.

\begin{definition}
\rm

$A$ is called a logical consequence of $\Gamma$
and is written as  $\Gamma \models A$,
if and only if for every {\bf M}, if ${\bf M} \models \Gamma$, then
${\bf M} \models A$ holds.

\end{definition}

\section{The User's rejections}

As we mentioned before, the users reject a specification when they
have found its counter example. In the first order logic, the user's
rejection can be defined by the models.

\begin{definition}
 \rm
User's rejection

Let $\Gamma \models A$. A model ${\bf M}$ is called a
user's rejection of $A$  if and only if ${\bf M} \models \neg A$. Let
\[
\Gamma_{M(A)}\equiv \{ A_{i} \mid A_{i} \in
\Gamma, \hspace{3mm} {\bf M} \models A_{i}, \hspace{3mm}
{\bf M} \models \neg A \}
\]

{\bf M} is called an {\em ideal} user's rejection of $A$ if and only if
$\Gamma_{M(A)}$ is {\em maximal} in the sense that there does not
exist another user's rejection ${\bf M'}$ of $A$,
such that $\Gamma_{M(A)} \subset \Gamma_{M'(A)}$.

\end{definition}

The above definition describe the following situation that $\Gamma
\vdash A$, but the users or designers have found a counter
example ${\bf M}$ that makes $\neg A$ true. $\Gamma_{M(A)}$ is a
subset of $\Gamma$ which does not contradict to $\neg A$. The
user's rejection meets the intuition that whether a specification
is accepted, depends only on whether its logical consequences
agree with user's requirements. The ideal user's rejection meets
the Occam's razor, which says: {\em Entities are not to be
multiplied beyond necessity}[3]. Here, it means that if a logical
consequence deduced from a specification is rejected by the users,
then the maximal subsets of the specification which is consistent
with the user's rejection must be retained and are assumed to be
true in the current stage of the development of the specification,
but the rest of laws contained in the specification must be
removed because they lead to the user's rejection.

In the rest of the paper, we consider ideal user's rejections
only, and simply call them user's rejections. Sometimes, we even
say that $\neg A$ is a user's rejection of $\Gamma$, it means that
$\Gamma \vdash A$ and there is an ideal user's rejection {\bf M}
and ${\bf M} \models \neg A$.

\begin{definition}
\rm

 (Maximal contraction).

Let $\Gamma \vdash A$ and $\Lambda \subset \Gamma$.  $\Lambda$ is
called a maximal contraction of $\Gamma$ by $\neg A$ if it is a maximal
subset of $\Gamma$ and is consistent with $\neg A$.

\end{definition}

\begin{example}\

\rm

Let $\Gamma \equiv \{ A, A \supset B, B \supset C, E\supset F\} $.
It can be proved that $\Gamma \vdash C$ holds.
Let $\neg C $ be a user's rejection. It cab be verified that
there are three maximal contractions:

\[
\{ A, A \supset B, E\supset F \},~~
\{ A, B \supset C, E\supset F \},~~
\{ A \supset B, B \supset C, E\supset F \}.
\]

\end{example}

\begin{lemma}\

\rm

 If $\Gamma \vdash A$ and $\Lambda$ is a maximal
contraction of $\Gamma$ by $\neg A$, then
there exists  a user's rejection {\bf M} of
$\Gamma$ by $A$ and ${\bf M} \models \neg A$ holds.

\end{lemma}

\noindent {\bf Proof:} The proof is directly from the definition.


\section{ The R-calculus}

The purpose of this section is to build an inference system about
logical connectives to remove the laws which is not consistent
with a given user's rejection.
It is called R-calculus. For a given $\Gamma \vdash A$,
the R-calculus is used to derive all maximal contractions of $\Gamma$ by $\neg A$.
In fact, if $\Gamma$ is not consistent,

the R-calculus is still employed to derive all maximal subsets of $\Gamma$
that is consistent with $\neg A$.

 In order to define the calculus, for a formal theory $\Gamma$,
a concept called R-condition of $\Gamma$ is to be introduced.
The R-condition is a kind of mirror reflection of the concept of T-condition
 used in the forcing theory [8].

\begin{definition}
\rm
R-condition of $\Gamma$

Let $\Gamma$ be a specification and $\Delta$ be a finite consistent
set of atomic formulas and the negations of atomic formulas.
$\Delta$ is called an R-condition of $\Gamma$,
if and only if for every $A \in \Delta$, $\Gamma \vdash \neg A$ holds.

\end{definition}

\begin{lemma}\

Let $\Gamma$ be a specification and $\Delta$ be a R-condition of $\Gamma$.
If $A \in \Delta$, then $A$ is a user's rejection of $\Gamma$.

\end{lemma}

\noindent {\bf Proof.} The proof is directly from the definition.

\begin{definition}
\rm
R-configuration

\[
\Delta|\Gamma
\]
is called a R-configuration, if and only if  $\Gamma$ is
a specification and $\Delta$ is a R-condition of $\Gamma$.
The R-configuration
$\Delta|\Gamma$ is read as $\Delta$ overrides $\Gamma$.

\end{definition}

$\Delta$ and $\Gamma$ can be written as sequences, such as
$A,B,\Delta_1$ and $A,B, \Gamma'$, etc.
Let $Delta$ be $A_1, A_2,\cdots, A_n$.
According to the above definitions, the R-configuration
$\Delta|\Gamma$ implies that
$\Gamma \vdash \neg A_1 \land \neg A_2, \cdots, \land A_n$ holds.
Let its proof tree be denoted by {\bf T}.

\begin{definition}
\rm
R-transition

\[
\Delta\mid\Gamma \Longrightarrow \Delta'\mid\Gamma'
\]
is called a R-transition. It means that the configuration
$\Delta\mid\Gamma$ is transformed to $\Delta'\mid\Gamma'$.

 $\Longrightarrow^{*}$ denotes a sequence of the transitions.
 $*$ denotes finite times of transitions but also infinite
 times and 0 times of transitions. The following R-transition
\[
\Delta\mid A, \Gamma \Longrightarrow \Delta\mid\Gamma
\]
means that $\Delta\mid\Gamma, A$ is transformed to $\Delta\mid\Gamma$,
and $A$ is deleted during the transition.

\end{definition}

R-calculus contains four kinds of transformation rules.
They are structural rules, the R-axiom, the R-cut rule
and the rules of logical connectives. As mentioned before,
for the writing simplicity, the proof tree is omitted from
$\mapsto$ and $\Longrightarrow $.

\vspace{3mm}

\noindent {\bf Structural rules}

\begin{definition}\
\rm

Contraction

\[
\Delta \mid A, A, \Gamma \Longrightarrow \Delta \mid A, \Gamma
\hspace{10mm}
 A, A, \Delta \mid \Gamma  \Longrightarrow A, \Delta \mid \Gamma
\]

Exchange
\[
\Delta \mid A, B, \Gamma  \Longrightarrow \Delta \mid B, A, \Gamma
  \hspace{10mm}
  A, B, \Delta \mid \Gamma \Longrightarrow B, A, \Delta \mid \Gamma \]
\end{definition}

The contraction rules mean that the same formulas occurring on one side
can be contracted to one. The exchange rules say that
a formula can be moved from one position to another
within one side of a configuration.

\begin{definition}

\rm
R-axiom

\[
  A, \Delta \mid \neg A, \Gamma \Longrightarrow A, \Delta \mid \Gamma
\]

\end{definition}

The R-axiom means that if $A$ (the atomic formula or the negation of
atomic formula) occurs on the left hand side and  its negation $ \neg A $
occurs on the right hand side, then $ \neg A $ must be deleted.

\begin{definition}
\rm
R-cut rule
\[
 \frac{\Gamma_{1},A \vdash B~~~A \mapsto B~~~B,\Gamma_{2} \vdash C~~~\Delta \mid
 C,\Gamma_{2}\Longrightarrow \Delta \mid \Gamma_{2} }{ \Delta
 \mid \Gamma_1, A, \Gamma_2 \Longrightarrow \Delta \mid \Gamma_1,
\Gamma_2}
\]
\end{definition}

Where $\Gamma = \Gamma_1, A, \Gamma_2$ and $ \Delta =\neg C,
\Delta'$. The R-cut means that $C$ is an atomic formula or the
negation of an atomic formula, and $C$ is not consistent with
$\Delta$. Furthermore, $B$ is a lemma used in the proof of $C$,
$A$ is contained in $\Gamma$ and is the necessary premise of
$B$. In this circumstance, $A$ must be eliminated.

\vspace{3mm}

\noindent {\bf Logical rules}

\vspace{2mm}

\begin{definition}
\rm
R-$\land$ rule

\[
\frac{\Delta \mid A, \Gamma \Longrightarrow \Delta \mid \Gamma}{
\Delta \mid A \land B, \Gamma \Longrightarrow \Delta \mid \Gamma}
\hspace{1cm} \frac{\Delta \mid B, \Gamma \Longrightarrow \Delta
\mid \Gamma}{ \Delta\mid A \land B, \Gamma \Longrightarrow \Delta
\mid \Gamma}
\]
\end{definition}

$A$ occurring in the numerator of the R-$\land$ rule means
that $\Delta \vdash \neg A$ holds. According to the $\land$ rule of
{\bf G} system, $\Delta \vdash \neg A \lor \neg B$ holds. That is
$\Delta \vdash \neg (A \land B)$ holds. Therefore, if $A$ is deleted,
then $A \land B$ must be deleted. Similarly, for the rule on the right,
if $B$ is deleted, then $A \land B$ must be deleted.

\begin{definition}
\rm

R-$\lor$ rule

\[
\frac{\Delta \mid A, \Gamma \Longrightarrow \Delta \mid \Gamma \hspace{1cm}
\Delta \mid B, \Gamma \Longrightarrow \Delta \mid \Gamma }{\Delta
\mid A \lor B, \Gamma \Longrightarrow \Delta \mid \Gamma }
\]

\end{definition}

Since $A$ and $B$ occurring in the numerator of R-$\lor$ rule
are going to be deleted, $\Delta \vdash \neg A$ and $\Delta \vdash
\neg B$ hold. According to the $\lor$ rule of the {\bf G} system,
$\Delta \vdash \neg A \land \neg B$ holds. The later implies $\Delta \vdash
\neg (A \lor B)$. Therefore, $A \lor B$ must be deleted.

\begin{definition}\
\rm
R-$\supset$ rule

\[
\frac{\Delta \mid \neg A, \Gamma \Longrightarrow \Delta \mid
\Gamma \hspace{1cm} \Delta \mid B, \Gamma \Longrightarrow \Gamma}{
\Delta \mid A \supset B, \Gamma \Longrightarrow \Delta \mid
\Gamma}
\]
\end{definition}

The R-$\supset$ rule holds since $(A \supset B ) \equiv (\neg A
\lor B)$.

\begin{definition}
\rm

R-$\forall$ rule

\[
\frac{\Delta \mid  A[t/x], \Gamma \Longrightarrow \Delta \mid \Gamma}{\Delta
\mid  \forall x A, \Gamma \Longrightarrow \Delta \mid \Gamma}
\]

where $t$ is a term and is free in $A$ for $x$.
\end{definition}

Since $ A[t/x]$ occurring in the numerator of the R-$\forall$ rule
is to be deleted, $\Delta \vdash \neg A[t/x]$ holds. It implies $\Delta
\vdash \neg \forall x A[x]$. Thus, $\forall x A[x]$ must be deleted.
R-$\forall$ means that if $ A[t/x]$ is not consistent with
$\Delta$, then $\forall x A(x)$ can not be consistent with $\Delta$.

\begin{definition}
\rm
R-$\exists$ rule

\[
\frac{\Delta \mid  A[y/x], \Gamma \Longrightarrow \Delta \mid
\Gamma}{\Delta \mid
 \exists x A, \Gamma \Longrightarrow \Delta \mid \Gamma}
\]
$y$ is an eigenvariable and it does not occur in the denominator
of the rule.

\end{definition}

If $A[y/x]$ is to be deleted in the numerator of the R-$\exists$ rule, then
$\Delta \vdash \neg A[y/x]$. According to the $\exists$ rule of the
{\bf G} system, $\Delta \vdash \neg \exists x A(x)$ holds.
Therefore,  $\exists x A(x)$ must be deleted.
This rule means that for any eigenvariable $y$,
if $ A[y/x]$ is not consistent with
$\Delta$, then $\exists x A(x)$ is not consistent with $\Delta$.

\begin{definition}
\rm
R-$\neg$ rule

\[
 \Delta \mid A, \Gamma  \Longrightarrow  \Delta \mid A', \Gamma
\]

 $A$ and $A'$ are defined as below:

\vspace{3mm}

\begin{tabular}{|c|c|c|c|c|c|c|} \hline
$A$ & $\neg (B \land C)$ & $\neg (B \lor C)$ & $\neg \neg B$ &
$\neg (B \supset C)$ & $\neg \forall x. B$ & $\neg \exists x. B$
\\ \hline $A'$ & $\neg B \lor \neg C$ & $\neg B \land \neg C$ &
$B$ & $ B \land \neg C $ & $ \exists x. \neg B$ & $\forall x. \neg
B$ \\\hline
\end{tabular}
\end{definition}

R-$\neg$  rule is an expansion rule. $\neg A$ occurring on the
left of the long right arrow is substituted by its equivalent
$A'$, and the "$\neg$" goes to the next level.

\begin{definition}
\rm

R-calculus

R-calculus is the set which consists of the structural rules, the
R-axioms, the R-cut rule, the R-$\land$ rule, the R-$\lor$ rule,
the R-$\supset$ rule, the R-$\forall$ rule, the R-$\exists$ rule,
and the R-$\neg$ rule.

An R-configuration $\Delta \mid \Gamma$
is called an
R-termination if there does not exist an R-rule
that can be applied to $\Delta \mid \Gamma$
with the exception of the structural rules.

\end{definition}

In summary, every R-configuration $\Delta | \Gamma $ consists two
parts: the left part $\Delta$ is a finite consistent set of atomic
formulas and the negations of atomic formulas, the right part
$\Gamma$ is a  finite set of sentences which may not be
consistent. For every $A \in \Delta$, $A$ is a user's rejection of
$\Gamma$. The R-calculus is an inference system. It can be used to
eliminate those laws which are not consistent with

　　　　$\Delta$.
The principles of eliminating are as below: The
law $A$ of $\Gamma$ (on the right hand side of $|$) is
to be eliminate if its negation $ \neg A$ occurs in $\Delta$ (on
the left hand side of $|$). If $A$ of $\Gamma$ is a
compound sentence, then whether $A$ is to be eliminated depends on
the eliminations of the components of $A$ and the meaning of the
logical connective occurring in $A$. The rule for a logical
connective of R-calculus is a mirror reflection of the rule for
the same logical connective of the first order inference system.

\section{Some Examples}

The following three examples are given to show how the R-calculus
can be used to delete the laws of $\Gamma$ which is not consistent
with its user's rejection.

\begin{example}\

\rm

Let $\Gamma \equiv\{ \forall x. A(x), \Gamma'\}$ and
$\Delta \equiv \{\neg A[c]\}$ holds. The latter means that we must accept
$\neg A[c]$, where $c$ is a constant. According to the R-axiom,

\[
\neg A[c]~ |~  A[c], \Gamma' \Longrightarrow
 \neg A[c]~ |~ \Gamma'
\]
holds. By the R-$\forall$ rule,
\[
\frac{ \neg A[c]~ |~  A[c], \Gamma' \Longrightarrow
 \neg A[c]~ |~ \Gamma'}{ \neg A[c]~|~ \forall x A(x),  \Gamma'
 \Longrightarrow \neg A[c]~ |~ \Gamma'}
\]
holds.  Thus, it is proved by the R-calculus that
$\forall x A(x)$ is not consistent with $\neg A[c]$.
and it should be eliminated from $\Gamma$. \hfill $\Box$

\end{example}

The following example demonstrates how to use the R-cut rule.

\begin{example}
\rm

Consider the example given in the beginning of the paper. Let

$$
\Gamma \equiv \{ A, A \supset B, B \supset C, E\supset F\}
$$

$\Gamma \vdash C$ holds. Suppose $\neg C$ is a user's rejection.
According to the definition, there exists three maximal
contraction of $\Gamma$ by $\neg C$:
\[
 \{ A, A \supset B, E\supset F \}, \hspace{5mm}
\{ A, B \supset C, E\supset F \} \hspace{5mm}
\{ A \supset B, B \supset C, E\supset F \}.
\]

In fact, each one of the above three can be derived by the R-calculus.
Consider $\{ A, A \supset B, E\supset F \}$ first. Let
\[
\Gamma_1   \equiv  \{ A, A \supset B\}~~~
\Gamma_2  \equiv  \{ E\supset F\}.
\]
By the {\bf G} system, both
\[
\Gamma_1,B\supset C \vdash  C~~\mbox{and}~~C,\Gamma_2 ~ \vdash C
\]
hold. According to the definition 2.1, $B\supset C$ is the necessary premise of $C$
and $B\supset C \mapsto C$ holds.
\[
\neg C \mid  C, \Gamma_2 \Longrightarrow \neg C \mid   \Gamma_2
\]
holds by the R-axiom. the R-cut rule is then applied and
\[
\neg C \mid \Gamma_1, B\supset C, \Gamma_2 \Longrightarrow
\neg C \mid \Gamma_1,\Gamma_2
\]
hold. Here $\Gamma_1,\Gamma_2$ is just $\{ A, A \supset B, E\supset F \}$.

Consider the second maximal contraction $\{ A, B \supset C, E\supset F \}$. Let

\[
\Gamma_1   \equiv  \{ A\}~~~
\Gamma_2  \equiv  \{ B \supset C,E\supset F\}.
\]
By the {\bf G} system,
\[
\Gamma_1,A\supset B~ \vdash ~ B~~\mbox{and}~~B,\Gamma_2 ~ \vdash C~~\mbox{hold.}
\]
Notice that $A\supset B$ is the premise of $B$ and
is in $\Gamma$. Thus, $A\supset B \mapsto B$ holds. According to the R-axiom,
\[
\neg C \mid C, \Gamma_2 \Longrightarrow \neg C \mid  \Gamma_2
\]
holds. Thus, the R-cut rule is applied and
\[
\neg C \mid \Gamma_1, A\supset B, \Gamma_2 \Longrightarrow
\neg C \mid \Gamma_1,\Gamma_2
\]
holds. Here $\Gamma_1,\Gamma_2$ is $\{ A, B \supset C, E\supset F \}$.
Finally, Let
\[
\Gamma_1   \equiv  \emptyset~~~
\Gamma_2  \equiv  \{ A\supset B, B \supset C,E\supset F\}.
\]

Using the similar proof strategy, the third maximal contraction
$\{ A \supset B, B \supset C, E\supset F \}$ can derived by
the R-calculus. \hfill $\Box$

\end{example}

In the above two examples, $\Gamma$ is a finite consistent set of laws.
The following example shows that if $\Gamma$ is not consistent,
the R-calculus still works to deduce all of maximal subset
of $\Gamma$ which is consistent with $\Delta$.
:

\begin{example}\

\rm

 Let $\Delta =\{x=x\}$ and $\Gamma=\{f(x)=y,~f(y)=z,~\neg
(f(f(x))=z)\}$. Obviously, $\Gamma$ is not consistent. But the R-cut
rule can be applied to deduce the maximal subsets which is consistent
with $\Delta$. For example, let $\Gamma_1=\{f(x)=y\}$,
 $\Gamma_2=\{\neg (f(f(x))=z)\}$.
First,
\[
\Gamma_1,~(f(y)=z) \vdash f(f(x))=z~~\mbox{and}~~(f(f(x))=z),
~\Gamma_2 \vdash \neg (x=x)
\]
holds. It is proved that $(f(y)=z)$ is the necessary premise of
$f(f(x))=z$. According to the R-axiom,
\[
(x=x)~| \neg (x=x) \Longrightarrow (x=x) | \emptyset
\]
holds. Therefore, by the R-cut rule,
\[
(x=x)|~\{f(x)=y,~f(y)=z,~\neg (f(f(x))=z)\} \Longrightarrow
(x=x)|~\{f(x)=y,~\neg (f(f(x))=z)\}
\]
holds.  It can be verified that $\{f(x)=y,~\neg (f(f(x))=z)\}$ is a maximal subset of $\Gamma$ and is consistent with $x=x$.
\hfill $\Box$

\end{example}

\section{ The Reachability and Completeness }

>From the examples given in the last section, we have found that
for the given $\Gamma$ and $\Delta$, every maximal contraction of
$\Gamma$ by $\Delta$ can be deduced by the R-calculus. This fact
is called the reachability of the R-calculus.

\begin{definition}
\rm

R-reachabilty

Let $\Delta \mid \Gamma$ be any given R-configuration, $\Gamma$ be
a specification, and $\Delta$ be an R-condition of $\Gamma$. The
R-calculus is reachable, if and only if for any given maximal
contraction $\Gamma'$ of $\Gamma$ by $\Delta$, there exists an
R-transition sequence such that
\[
\Delta \mid \Gamma \Longrightarrow^{*} \Delta \mid \Gamma'
\]
holds, where $\Delta \mid \Gamma'$ is an R-termination.

\end{definition}

\begin{lemma}\
\rm

Let $\Delta \mid \Gamma$ be a given R-configuration,
$\Gamma$ be a specification, and $\Delta$ be
an R-condition of $\Gamma$.
If $\Gamma_1$ is a maximal contraction of $\Gamma$ by $\Delta$,
then there exists a sequence of R-transitions, such that
\[
\Delta \mid \Gamma \Longrightarrow^{*} \Delta \mid \Gamma_1
\]
holds.

\end{lemma}

\noindent {\bf Proof.} Consider the simple case that $\Delta$
contains only one element, $\Delta \equiv \{ \neg A \}$, and $A $
is an atomic formula or the negation of atomic formula.
$\Gamma \vdash A$ holds. Let
$\Gamma_1$ be a maximal contraction of $\Gamma$ by $\neg A $, and
let $\Gamma_2 \equiv \Gamma - \Gamma_1$.

The aim is to prove that for any $B \in \Gamma_2$, $B$ will be
eliminated by the R-calculus. To do so, let $\Gamma_3 \equiv
\Gamma_2-\{B\}$. Thus, $\Gamma=\Gamma_3,B,\Gamma_1$.
First, $\Gamma_3,B \vdash B$ holds. Second, $B$ on the left of
$\vdash$ is the necessary premise of $B$ on the right of
$\vdash$. Since $\Gamma_1$ is a maximal contraction
of $\Gamma$ by $\neg A$, $\Gamma_1,B \vdash A$ holds.
\[
\neg A \mid A, \Gamma_1 \Longrightarrow \neg A~\mid \Gamma_1
\]
is an application of the R-axiom. By the R-cut rule,

\[
\frac{\Gamma_3,B \vdash B~~~B \mapsto B~~~\Gamma_1,B \vdash A~~~
\neg A \mid A, \Gamma_1 \Longrightarrow \neg A~\mid \Gamma_1}
{\neg A \mid \Gamma_3,B, \Gamma_1 \Longrightarrow \neg A~\mid
\Gamma_3, \Gamma_1}
\]

Thus, $B$ is eliminated. Therefore, every law
of $\Gamma_2$ should be eliminated by the R-calculus.
 \hfill $\Box$

The converse of the lemma is not true. For every sequence of R-transitions:
\[
\Delta \mid\Gamma \Longrightarrow^{*} \Delta' \mid \Gamma'
\]
where $\Delta' \mid \Gamma'$ is an R-termination, $\Gamma'$ may
not be a maximal contraction of $\Gamma$ by $\Delta$. Consider the
following example:

\begin{example}
\rm
Let
$$
\Gamma \equiv \{ A, A \supset B, B \supset C, A \supset E, E\supset C\}
$$

$\Gamma \vdash C$ holds. Suppose that $C$ is rejected by the
users. Using the R-cut rule, We can eliminate $A\supset B$. And
then, since
$$
 A,  A \supset E, E\supset C \vdash C
$$
we apply the R-cut rule again and eliminate $A$. Thus, we have:
$$
\{ B \supset C, A \supset E, E\supset C \}.
$$
The above set is not a maximal contraction of $\Gamma$ by $\neg
C$. The maximal contraction is
\[
\{ A \supset B, B \supset C, A \supset E, E\supset C \}.
\]
\hfill $\Box$

\end{example}

\begin{lemma}\
\rm

Let $A~ |~ \Gamma$ be an R-configuration, $A$ be an atomic formula
or the negation of an atomic formula. If $\Gamma$ is consistent
with $A$, then $A | \Gamma$ is an R-terminated configuration.

\end{lemma}

\noindent {\bf Proof:} Since $\Gamma$ is finite, $\Gamma$ can be
written as $A_1 \land \cdots \land A_n$. Let $r(\Gamma)$ be the
rank of $\Gamma$ [1]. The proof is given by induction on $r(A_1
\land \cdots \land A_n)$ as below:

If $r(\Gamma)=1$, $\Gamma$ is an atomic formula. It can not be
eliminated since it is consistent with $A$. By the definition, $A
| \Gamma$ is R-termination.

Suppose that the lemma holds for  $r(\Gamma)<k$. Consider the case
of $r(\Gamma)=k$. Let $\Gamma$ be $B,\Gamma'$, where
$r(\Gamma')<k$, and $\Gamma'$ is consistent with $A$. $B$ can be
one of the following cases:

\begin{enumerate}

\item $B$ is an atomic formula. $B$ can not be $\neg A$ since
$\Gamma$ is consistent with $A$. Therefore, $A | \Gamma$ is an
R-termination.

\item $B$ is $B_1 \lor B_2$. According to the $R-\lor$ rule, $B$
is eliminated if and only if $B_1$ in $A | B_1,\Gamma'$ is to be
eliminated, and $B_2$ in $A | B_2,\Gamma'$ is also to be
eliminated. Since  $ B_1,\Gamma'$ with $A$, and $
r(B_1,\Gamma')<k$ holds. According to the inductive hypothesis,
$B_1$ in $A | B_1,\Gamma'$ can not be eliminated. Similarly, $B_2$
in $A | B_2,\Gamma'$ is also not eliminated. Therefore, $A |
\Gamma$ is an R-termination.

\item Similarly, we can prove the cases that $B$  is $B_1 \land
B_2$ and $B$ is $B_1 \supset B_2$.

\item $B$ is $\forall x B_1$. According to the $R-\forall$ rule,
$B_1$ is eliminated if and only if $B_1[t/x]$ in $A |
B_1[t/x],\Gamma'$ is to be eliminated. Since $\{
B_1[t/x],\Gamma'\}$ is consistent with $A$, And $
r(B_1[t/x],\Gamma')<k$ holds. According to the inductive premise,
$B_1[t/x]$ in $ B_1[t/x],\Gamma'$ can not be eliminated. Thus, $A
| \Gamma$ is an R-termination.

\item Similarly, we can prove the case of that  $B$ is $\exists x.
B_1$.

\item Finally, we prove that every one of $A_1,\cdots,A_n$ in
$\Gamma$ can not be eliminated by the R-cut rule. For each $A_k$,
$k=1,\cdots,n$, the R-cut rule is applied only in the circumstance
that there exists $B$, such that $A_1,\cdots,A_{k-1},A_k \vdash
B$, $A_k \mapsto B$, and $B, A_{k+1},\cdots, A_{n} \vdash P$
holds, and $P$ in $A | P$, and $A_{k+1},\cdots, A_{n}$ is to be
eliminated. Since $\Gamma$ is consistent with $A$,
$\{P,A_{k+1},\cdots, A_{n} \}$ is also consistent with $A$,
Furthermore, $P$ is an atomic formula or the negation of an atomic
formula. We know that $r(P, A_{k+1},\cdots, A_{n})\leq k$ holds.
According to the item 1,  we know that $A | P, A_{k+1},\cdots,
A_{n}$ is an R-termination. So $P$ can not be eliminated.
Therefore, the R-cut rule can not be applied. \hfill $\Box$
\end{enumerate}

\begin{theorem}\
\rm

The R-calculus is reachable.

\end{theorem}

\noindent {\bf Proof:} Let $\Delta \mid \Gamma$ be a given
 R-configuration, where $\Gamma$ is a finite set of sentences,
 and $\Delta$ is an R-condition of $\Gamma$.
 Consider the simple case that $\Delta$ contains only
 one element $A$. Let $\Gamma'$ be a maximal
 contraction of $\Gamma$ by $A$.
For every $B$ in $\Gamma -\Gamma'$, by the lemma 6.1, there exists
a sequence of R-transitions at the end of which $B$ is eliminated.
Since  $\Gamma -\Gamma'$ is a finite set of sentences, the above
the sequences can be concatenated to form a sequence of
R-transitions:
\[
A~ |~\Gamma \Longrightarrow^{*} A~|~\Gamma
\]
where $A~ |~\Gamma'$ is an R-termination by the lemma 6.2. \hfill
$\Box$

\begin{definition}
\rm

R-completeness

Let $\Delta \mid \Gamma$ be any R-configuration, where
$\Gamma$ is a specification and $\Delta$ is R-condition of $\Gamma$.

The R-calculus is R-complete, if and only if for a given
R-configuration $\Delta \mid \Gamma$ and every ideal user's
rejection {\bf M}, if ${\bf M} \models \Delta$, then there exists
a transition sequence:
$$
\Delta \mid\Gamma \Longrightarrow^{*} \Delta \mid \Gamma',
$$
where $\Delta \mid \Gamma'$ is a termination and
\[
\Gamma'=  \{ A ~~ | M  \models A~~ and ~~ A \in \Gamma \}
\]
holds. \hfill $\Box$

\end{definition}

\begin{theorem}\

\rm

 The R-calculus is R-complete.

\end{theorem}

\noindent {\bf Proof:} This theorem is a corollary of the lemma
3.1 and the theorem  of R-reachability. \hfill $\Box$

\section{Related works}

In 1985, G\"ardenfors and his colleagues introduced their theory
of changes [4]. The theory addresses the proof-theoretic concepts
of the expansion, the contraction, and the revision in the scope
of propositional logic. The maximal contraction given here can be
viewed a special kind of AGM's contraction, but in the scope of
the first order logic. The user's rejection is a corresponding
model-theoretic concept of the maximal contractions [6,9].

The AGM's theory focuses on building the systems of the
propositions of the expansion, the contraction and the revision,
and on studying the properties of these systems [4,5]. The
principal difference between the AGM's theory and The R-calculus
is as the following: the aim of designing the R-calculus is to
build a transition system that can deduce all maximal contractions
from a given formal theory $\Gamma$ and its user's rejection $A$.

Finally, it is believed that using the methods given in [2],
certain proper type theories based on the R-calculus can be
constructed and the corresponding interactive tools can be
implemented to develop the specifications .

\vspace{5mm}

\noindent { \bf Acknowledgement:}  The author would like to take
this chance to thank Dr. Zhang Yuping. His counter examples helped
the author to find the current version of the definition of the
necessary premise.

\vspace{5mm}

\noindent { \bf  References}

\smallskip

\begin{description}

\item{[1]} Gallier, J.H., Logic for Computer
  Science, foundations of automatic theorem proving.
John  Wiley \& Sons, 1987, 147-158, 162-163, 197-217.

\item{[2]} Paulson, L., Logic and Computations,
 Cambridge University Press, 1987, 38-50.

\item{[3]} Flew, A., A Dictionary of Philosophy,
Pan Books Ltd, 1979.

\item{[4]} Alchourr\'on, C.E., G\"ardenfors, R. and Makinson, D.,
On the logic of theory change: partial meet contraction and revision
functions, The Journal of Symbolic
Logic, Vol.50, No.2, June, 1985.

\item{[5]} G\"ardenfors, P., Knowledge in Flux,
The MIT Press, 1988.

\item{[6]} Li, W., An Open Logic System, Science in China, Series A, March, 1993.

\item{[7]} Shoenfield, J.R., Mathematical Logic, Addison-Wesley,
Reading, Mass, 1967, 74-75.

\item{[8]} Forcing, Burgess,  Handbook of mathematical logic,
edited by J., Barwise,  North-Holland Publishing Company, 1977.

\item{[9]} Li W., A Logical Framework for the evolution of
Specifications. ESOP'94, LNCS 788, 1994, Springer-Verlag.

\item{[10]} Burstall, R. and Goguen J.A., Putting
theories together to make specifications, Proc. 5th. IJCAI
Cambridge, Mass, 1045-1058 (1977).

\item{[11]} Bj{\o}rner, D., and Jones, C., Formal
specification and Software Development, Prentice Hall
International, 1983.

\item{[12]} Nordstr\"{o}m, B., and Smith, J.,
 Propositions and specifications of programs in
Martin-L\"{o}f's type theory, BIT 24 (1984),
pp 288-301.

\item{[13]} D. Sannella and A. Tarlecki,
Toward formal development of programs from algebraic
specifications: implementations revisited.
Acta Informatica 25,233-281(1988).

\end{description}

\end{document}